\documentclass[15pt a4paper]{article}

\usepackage{latexsym}
\usepackage{amsmath}
\usepackage{amssymb}
\usepackage{graphicx}
\graphicspath{{./}{./figs/}}

\begin{document}

\title{Hopf Bifurcation and Chaos in a Single Inertial Neuron Model with Time Delay \thanks{European Physical Journal B, Vol. 41, pp. 337-343, 2004.}}
\author{Chunguang Li$^1$\thanks{Corresponding author, Email: cgli@uestc.edu.cn}, Guanrong Chen$^2$, Xiaofeng Liao$^1$, Juebang
Yu$^1$}

\date{\small $^1$Institute of Electronic Systems, School of Electronic
Engineering, \\University of Electronic Science and
Technology of China,\\ Chengdu, 610054, P. R. China.\\
$^2$Department of Electronic Engineering, City University of Hong
Kong, \\83 Tat Chee Avenue, Kowloon, Hong Kong, P. R. China.}
\maketitle

\begin{abstract}
A delayed differential equation modelling a single neuron with
inertial term is considered in this paper. Hopf bifurcation is
studied by using the normal form theory of retarded functional
differential equations. When adopting a nonmonotonic activation
function, chaotic behavior is observed. Phase plots, waveform
plots, and power spectra are presented to confirm  the chaoticity.

Keywords: Hopf bifurcation, time delay, neural network, chaos
\end{abstract}

\section{Introduction}

In recent years, dynamical characteristics of neural networks have
become a focal subject of intensive research studies. Bifurcations
and chaotic phenomena have been investigated in various neural
networks. For example, chaotic solutions were obtained in a neural
network consisting of 26 neurons in [1]. Numerical solutions of
differential equations with electronic circuit models of chaotic
neural networks were qualitatively studied in [2]. In [3], a
chaotic neural network with four neurons was investigated. Chaotic
behavior was found in a cellular neural network with three cells
in [4]. In [5], chaotic phenomenon in a three-neuron hysteretic
Hopfield-type neural network was discussed. In [6], a
high-dimensional chaotic neural network under external sinusoidal
force was studied. In [7], bifurcation and chaos as well as their
control in a system of strongly connected neural oscillators were
discussed. In [8, 9], a discrete-time transiently chaotic neural
network was studied. The chaotic phenomenon in a neural network
learning algorithm was reported in [10]. Moreover, chaotic
behaviors of inertial neural networks are studied in [11, 12]. On
the other hand, there are extensive literatures studied neural
network models with delays. For example, bifurcations and chaotic
dynamics of neural networks with discrete and distributed delays
were studied in [13-21].

In this paper, the dynamical behaviors of a single delayed neuron
model with inertial terms are investigated. The work presented in
this paper can be considered as an extension of the works for
inertial neural network without delays [11, 12] to the case with
delays, or an extension of the work for single neuron without
inertial terms [13] to the case with inertial terms.

The paper is organized as follows. The delayed inertial neuron
model is described, and the local stability and the existence of
Hopf bifurcation is studied in Section 2. In Section 3, the
properties of the bifurcating periodic solutions are analyzed
based on the normal form theory developed in [22]. To justify the
theoretical analysis, a numerical example is given in Section 4.
In Section 5, the observed chaotic behavior of the model with a
nonmonotonic activation function is reported. Finally, conclusions
are drawn in Section 6.

\section{Local stability and the existence of Hopf bifurcation}
The single inertial neuron with time delay, similar to that in
[13] but with an inertial term, is described by
\begin{equation}
\ddot{x}=-a\dot{x}-bx+cf(x-hx(t-\tau))
\end{equation}
where constants $a,b,c>0, h\geq 0$, and $\tau > 0$ is the time
delay. Without loss of generality, assume that the activation
function $f(.)$ in the above equation is a nonlinear function and
its third-order continuous derivative exists. Define
\begin{equation*}
x_1(t)\equiv x(t)-hx(t-\tau),\hspace{0.5cm} t\in[-\tau, +\infty)
\end{equation*}
Then, we have
\begin{equation}
\begin{array}{rl}
\dot{x}_1(t)&=x_2(t)\\
\dot{x}_2(t)&=-ax_2(t)-bx_1(t)+cf(x_1(t))-chf(x_1(t-\tau))
\end{array}
\end{equation}
The phase space is $C:=C([-\tau,0];R^2)$. Throughout this paper,
assume that the following conditions are satisfied:
\begin{equation*}
f(0)=0,\,f'(0)>0, \,|c(1-h)|<1/f'(0)
\end{equation*}
It is clear that (2) has an unique equilibrium (0, 0) under the
above condition. It is also easy to see that if there is no delay
term in (1), i.e. $h=0$, then the model is asymptotically stable
when
\begin{equation}
b-cf'(0)>0.
\end{equation}
In the following, we estimate the value of $h$ that preserves the
system stability under the above condition.

For convenience, we restate here a result of Bellman and
Cooke [24, Theorem 13.9]. \\
{\bf Lemma 1} [24]: Let $H(z)\equiv(z^2+pz+q)e^z+r$, where $p$ is
real and positive, $q$ is real and nonnegative, and $r$ is real.
Denote by $a_k\, (k\geq 0)$ the sole root of the equation
\begin{equation*}
\mbox{cot}\,a=(a^2-q)/ap
\end{equation*}
which lies on the interval $(k\pi,k\pi+\pi)$. Define the natural
number $n$ as follows:
\begin{enumerate}
\item if $r\geq 0$ and $p^2\geq 2q$, $n=1$; \item if $r\geq 0$ and
$p^2< 2q$, $n$ is the odd integer $k$ for which $a_k$ lies closest
to $\sqrt{q-p^2/2}$; \item if $r<0$ and $p^2\geq 2q$, $n=2$; \item
if $r< 0$ and $p^2< 2q$, $n$ is the even integer $k$ for which
$a_k$ lies closest to $\sqrt{q-p^2/2}$.
\end{enumerate}
Then, a necessary and sufficient condition under which all the
roots of $H(z)=0$ lie to the left of the imaginary axis is that
\begin{enumerate}
\item $r\geq 0$ and $(r\,\mbox{sin}\,a_n)/pa_n<1$, or \item
$-q<r<0$ and $(r\,\mbox{sin}\,a_n)/pa_n<1$.
\end{enumerate}

Separating the linear and the nonlinear terms, (2) becomes
\begin{equation}
\dot{x}=L(x_t)+F(x_t)
\end{equation}
where $x_t\in C,x_t(\theta)=x(t+\theta),-\tau \leq\theta\leq 0$,
and $L:C\rightarrow R^2,\,F:C\rightarrow R^2$ are given
respectively by
\begin{equation}
\begin{array}{rl}
L(\varphi)&=\left[\begin{array}{c}\varphi_2(0)\\-(b-cf'(0))\varphi_1(0)
-a\varphi_2(0)-chf'(0)\varphi_1(-\tau)\end{array}\right]\\
F(\varphi)&=\left[\begin{array}{c}0\\\frac{cf''(0)}{2}\varphi_1^2(0)
+\frac{cf'''(0)}{6}\varphi_1^3(0)-\frac{chf''(0)}{2}\varphi_1^2(-\tau)
-\frac{chf'''(0)}{6}\varphi_1^3(-\tau)+\cdots\end{array}\right]
\end{array}
\end{equation}
with $\varphi=(\varphi_1,\varphi_2)\in C$. Here and throughout
this paper, we refer to [25] for notation and classical results on
functional differential equations (FDEs), including such as Eq.
(4).

The characteristic equation for the linearization of Eq.(4) at (0,
0) is
\begin{equation}
\lambda^2+a\lambda+(b-cf'(0))+chf'(0)e^{-\lambda\tau}=0
\end{equation}
Let $s=\lambda\tau$. Then we have
\begin{equation}
[s^2+a\tau s+(b-cf'(0))\tau^2]e^s+chf'(0)\tau^2=0
\end{equation}

The fixed point is locally stable if all roots of the above
equation have negative real parts [25]. For each $\tau$, we are
interested in the maximum value of $h$ such that the system is
locally stable.\\
{\bf Theorem 1}: Denote by $w_k\, (k\geq 0)$ the sole root of the
equation
\begin{equation*}
\mbox{cot}\,w=[w^2-(b-cf'(0))\tau^2]/a\tau w
\end{equation*}
which lies on the interval $(k\pi,k\pi+\pi)$. Define the nature
number $n$ as follows:
\begin{enumerate}
\item if $a^2\geq 2(b-cf'(0))$, $n=1$; \item if $a^2 <
2(b-cf'(0))$, $n$ is the odd $k$ for which $w_k$ lies closest to
$\sqrt{(b-cf'(0))-a^2/2}\,\tau$
\end{enumerate}
Then, under condition (3), a necessary and sufficient condition
that the solution of (2) is asymptotically stable is that
\begin{equation*}
h<\frac{aw_n}{cf'(0)\tau\mbox{sin}w_n}
\end{equation*}
{\bf Proof}: Since $a>0,\tau>0,f'(0)>0, b-cf'(0)>0,h\geq 0$, a
direct application of Lemma 1 to (7) with $p=a\tau,
q=(b-cf'(0))\tau^2$ and $r=chf'(0)\tau^2$ proves the claim.

In the following, we study the existence of Hopf bifurcation in
Eq. (2) by choosing $h$ as the bifurcation parameter. First, we
would like to know when Eq. (6) has purely imaginary roots
$\lambda=\pm i\omega_0\,(\omega_0>0)$ at $h=h_0$. If $\lambda=\pm
i\omega_0,\,\omega_0>0$, we have
\begin{equation*}
\begin{array}{rl}
chf'(0)\mbox{cos}\,\omega_0\tau &=\omega_0^2-(b-cf'(0))\\
chf'(0)\mbox{sin}\,\omega_0\tau &=a\omega_0
\end{array}
\end{equation*}
The above equations imply that
\begin{equation*}
\mbox{cot}\,\omega_0\tau=\frac{\omega_0^2-(b-cf'(0))}{a\omega_0}
\equiv g(\omega_0)
\end{equation*}
and, consequently,
\begin{equation*}
g'(\omega)=\frac{\omega^2+(b-cf'(0))}{a\omega^2}>0
\end{equation*}
So, $g(\omega)$ is strictly monotonically increasing on $(0,
+\infty)$, with $\lim_{\omega\rightarrow 0}g(\omega)=-\infty$ and
$\lim_{\omega\rightarrow +\infty}g(\omega)=+\infty$. Clearly,
$g(\omega)$ intersects $\mbox{cot}\,\omega\tau$ only at a point.
Hence, $\lambda=\pm i\omega_0$ are simple roots of Eq. (6), where
$\omega_0$ is the unique root of $\mbox{cot}
\omega\tau=\frac{\omega^2-(b-cf'(0))}{a\omega}$, and
$h_0=\frac{a\omega_0}{cf'(0)\mbox{sin}\,\omega_0\tau}$.

From [26], we know that all the other roots of Eq. (6) have
negative real parts. We proceed to calculate $Re[d\lambda /dh]$ at
$h=h_0$. Differentiating Eq. (6) with respect to $h$ yields
\begin{equation*}
\frac{d\lambda}{dh}=\frac{cf'(0)e^{-\lambda\tau}}{ch\tau
f'(0)e^{-\lambda\tau}-2\lambda-a}
\end{equation*}
So, we have
\begin{equation*}
\mbox{Re}\left[\frac{d \lambda}{d
\,h}\right]_{\lambda=i\omega_0\atop h=h_0}=
\frac{cf'(0)[cf'(0)h_0\tau-a\mbox{cos}\,\omega_0\tau+2\omega_0\mbox{sin}\,\omega_0\tau]}
{(ch_0\tau f'(0)\mbox{cos}\,\omega_0\tau-a)^2+(ch_0\tau
f'(0)\mbox{sin}\,\omega_0 \tau+2\omega_0)^2}
\end{equation*}

From the above analysis, we have the following result.\\
{\bf Theorem 2}: Under condition (3), if $cf'(0)h_0 \tau-a
\mbox{cos}\, \omega_0\tau+2 \omega_0\mbox{sin}\,\omega_0\tau \neq
0$, then as $h$ pass through the critical value
$h_0=\frac{a\omega_0} {cf'(0)\mbox{sin}\,\omega_0\tau}$, there is
a Hopf bifurcation of system (1) at its equilibrium (0, 0), where
$\omega_0$ is the sole root of $\mbox{cot}\, \omega\tau=
\frac{\omega^2-(b-cf'(0))}{a\omega}$.

{\bf Remark}: Note that if we let $w = \omega_0 \tau$, then the
constant $h_0$ in Theorem 2 can be rewritten as
$h_0=\frac{aw}{cf'(0) \tau \mbox{sin} w}$, which is consistent to
that in Theorem 1.

\section{Direction and stability of bifurcating periodic solutions}
In this section, we study the direction and stability of the
bifurcating periodic solutions. The method used here is based on
the normal form theory developed by Faria and Magalh\~aes [22].
This method computes normal forms for retarded functional
differential equations, without computing beforehand the center
manifold of the singularity.

As in [21], in the following we assume $f''(0)=0, f'''(0)\neq 0$.
Define $\Lambda=\{-i\omega_0,i\omega_0\}$ and introduce the new
parameter $\beta=h-h_0$. Eq. (4) can be rewritten as
\begin{equation}
\dot{x}=L_0(x_t)+F_0(x_t,\beta)
\end{equation}
where
\begin{equation*}
\begin{array}{rl}
L_0(\varphi)&=\left[\begin{array}{c}\varphi_2(0)\\-(b-cf'(0))\varphi_1(0)
-a\varphi_2(0)-ch_0f'(0)\varphi_1(-\tau)\end{array}\right]\\
F_0(\varphi)&=\left[\begin{array}{c}0\\
-cf'(0)\beta\varphi_1(-\tau) +\frac{cf'''(0)}{6}\varphi_1^3(0)
-\frac{cf'''(0)}{6}(h_0+\beta)\varphi_1^3(-\tau)+\cdots\end{array}\right]
\end{array}
\end{equation*}
Following the formal adjoint theory of FDEs [25], let the phase
space $C$ be decomposed according to $\Lambda$ as $C=P\bigoplus
Q$, where $P$ is the center space for $\dot{x}=L_0(x_t)$, i.e.,
$P$ is the generalized eigenspace associated with $\Lambda$.
Consider the bilinear form $(\cdot,\cdot)$ associated with the
linear equation $\dot{x}=L_0(x_t)$ [23]. Let $\Phi$ and $\Psi$ be
bases for $P$ and $P^*$ associated with the eigenvalues $\pm
i\omega_0$ of the adjoint equation, respectively, and normalize
them so that $(\Phi, \Psi)=I$. In complex coordinates, $\Phi,\Psi$
are written as $2\times 2$ matrices of the form
\begin{equation}
\begin{array}{l}
\Phi(\theta)=[\phi_1(\theta),\phi_2(\theta)],\hspace{0.2cm}\phi_1(\theta)=e^{i\omega_0\theta}v,
\hspace{0.2cm} \phi_2(\theta)=\overline{\phi_1(\theta)},
\hspace{0.2cm} -\tau \leq\theta\leq 0,\\
\Psi(s)=\left[\begin{array}{c}\psi_1(s)\\
\psi_2(s)\end{array}\right], \hspace{0.2cm}
\psi_1(s)=e^{-i\omega_0s}u^T, \hspace{0.2cm}
\psi_2(s)=\overline{\psi_1(s)},\hspace{0.2cm}0\leq s\leq \tau
\end{array}
\end{equation}
where the bar means complex conjugation, $u^T$ is the transpose of
$u$, and $u,v$ are vectors in $C^2$ such that
\begin{equation}
L_0(\phi_1)=i\omega_0v,\hspace{0.2cm} u^TL_0(e^{i\omega_0s} I)
=i\omega_0u^T, \hspace{0.2cm} (\psi_1,\phi_1)=1.
\end{equation}
Note that $\dot{\Phi}=\Phi B$, where $B$ is the diagonal matrix
$B=\mbox{diag}(i\omega_0,-i\omega_0)$. From (10), we have
\begin{equation}
\begin{array}{c}
v_2=i\omega_0v_1, \hspace{0.2cm}[-(b-cf'(0))-ch_0f'(0)e^{-i\omega_0\tau}]v_1=(a+i\omega_0)v_2, \\
u_1=(a+i\omega_0)u_2,\hspace{0.2cm}[-(b-cf'(0))-ch_0f'(0)e^{-i\omega_0\tau}]u_2
= i\omega_0u_1,
\end{array}
\end{equation}
Hence, we can select
\begin{equation}
v=\left[\begin{array}{c}v_1\\v_2\end{array}\right]=\left[\begin{array}{c}1\\i\omega_0\end{array}\right],
\hspace{0.5cm}
u=\left[\begin{array}{c}u_1\\u_2\end{array}\right]=u_1\left[\begin{array}{c}1\\
\frac {1}{a+i\omega_0}\end{array}\right]
\end{equation}
with
\begin{equation*}
u_1=\frac{a+i\omega_0}{a+2i\omega_0-ch_0f'(0)\tau
e^{-i\omega_0\tau}}
\end{equation*}

Here and in the following, we refer to [22] for results and
explanations of several notations involved. Enlarging the phase
space $C$ by considering the space $BC$ and using the
decomposition $x_t=\Phi z(t)+y_t, z\in C^2,y_t\in Q^1$, we
decompose system (8) as
\begin{equation}
\left\{ \begin{array}{l}\dot{z}=Bz+\Psi(0)F_0(\Phi z+y,\beta)\\
\dot{y}=A_{Q^1}y+(I-\pi)X_0F_0(\Phi z+y,\beta)\end{array}\right.
\end{equation}
Consider the Taylor formula
\begin{equation*}
\Psi(0)F_0(\Phi z+y,\beta)=\frac{1}{2}f_2^1(z,y,\beta)+
\frac{1}{6}f_3^1(z,y,\beta)+h.o.t.
\end{equation*}
where $f_j^1(z,y,\beta)\,(j=2,3)$ are homogeneous polynomials in
$(z,y,\beta)$ of degree $j$ with coefficients in $C^2$ and
$h.o.t.$ stands for higher order terms. The normal form on the
2-dimensional center manifold of the origin and $\beta=0$ is given
by
\begin{equation}
\dot{z}=Bz+\frac{1}{2}g_2^1(z,0,\beta)+\frac{1}{6}g_3^1(z,0,\beta)+h.o.t.
\end{equation}
where $g_2^1,g_3^1$ are second and third order terms in
$(z,\beta)$, respectively.

Using the notations in [22], we have
\begin{equation*}
\frac{1}{2}g_2^1(z,0,\beta)=\frac{1}{2}\mbox{Proj}_{\mbox{Ker}(M_2^1)}f_2^1(z,0,\beta)
\end{equation*}
where $\mbox{Proj}_{S}f$ is the projection of $f$ on $S$, and
\begin{equation*}
\mbox{Ker}(M_2^1)=\mbox{span}\left\{\left(\begin{array}{c}z_1\beta\\0\end{array}\right),
\left(\begin{array}{c}0\\z_2\beta\end{array}\right)\right\}
\end{equation*}
After some computation, we obtain
\begin{equation*}
\frac{1}{2}g_2^1(z,0,\beta)=\left[\begin{array}{c}A_1z_1\beta\\
\overline{A_1}z_2\beta\end{array}\right]
\end{equation*}
with
\begin{equation*}
A_1=-cf'(0)e^{-i\omega_0\tau}u_2
\end{equation*}

To compute the cubic terms, we first deduce that, after the change
of variables that transformed the quadratic terms
$f_2^1(z,y,\beta)$ into $g_2^1(z,y,\beta)$, the coefficients of
the third order terms at $y=0,\beta=0$ are still given by
$\frac{1}{6}f_3^1(z,0,0)$ (because $f''(0)=0$, implying
$f_2^1(z,y,0)=0$). This implies that [22]
\begin{equation*}
\frac{1}{6}g_3^1(z,0,\beta)=\frac{1}{6}\mbox{Proj}_{\mbox{Ker}(M_3^1)}f_3^1(z,0,\beta)
\end{equation*}
where
\begin{equation*}
\mbox{Ker}(M_3^1)=\mbox{span}\left\{\left(\begin{array}{c}z_1^2z_2\\0\end{array}\right),
\left(\begin{array}{c}z_1\beta^2\\0\end{array}\right),\left(\begin{array}{c}0\\z_1z_2^2\end{array}\right),
\left(\begin{array}{c}0\\z_2\beta^2\end{array}\right)\right\}
\end{equation*}
However, the terms $O(|z|\beta^2)$ are irrelevant to the
determination of the generic Hopf bifurcation. Hence, we write
\begin{equation*}
\frac{1}{6}g_3^1(z,0,\beta)=\frac{1}{6}\mbox{Proj}_Sf_3^1(z,0,0)+O(|z|\beta^2)
\end{equation*}
for
\begin{equation*}
S:=\mbox{span}\left\{\left(\begin{array}{c}z_1^2z_2\\0\end{array}\right),
\left(\begin{array}{c}0\\z_1z_2^2\end{array}\right)\right\}
\end{equation*}
Some computations yield
\begin{equation*}
\frac{1}{6}g_3^1(z,0,\beta)=\left[\begin{array}{c}A_2z_1^2z_2\\
\overline{A_2}z_1z_2^2\end{array}\right]+O(|z|\beta^2)
\end{equation*}
with
\begin{equation*}
A_2=\frac{cf'''(0)}{2}(1-h_0e^{-i\omega_0\tau})u_2
\end{equation*}

Thus, we obtain the normal form (14) with the coefficients
$A_1,A_2$ given explicitly in terms of the original equation (4),
without the need to compute the center manifold beforehand. The
normal form (14) can be written in real coordinates $(x,y)$,
through the change of variables $z_1=x-iy,z_2=x+iy$. In polar
coordinates $(r,\theta), x=r\mbox{cos}\theta,
y=r\mbox{sin}\theta$, this normal form becomes
\begin{equation}
\left\{\begin{array}{l}\dot{r}=K_1\beta r+K_2r^3+O(\beta^2r+|(r,\beta)|^4)\\
\dot{\theta}=-\omega_0+O(|(r,\beta)|) \end{array} \right.
\end{equation}
where $K_1:=Re\,A_1,K_2:=Re\,A_2$.

We have the following theorem.\\
{\bf Theorem 3}: In formula (15), the sign of $K_1K_2$ determines
the direction of the Hopf bifurcation: if $K_1K_2<0\,(>0)$, then
the Hopf bifurcation is supercritical (subcritical) and the
bifurcating periodic solutions exist for $h>h_0\,(<h_0)$; the sign
of $K_2$ determines the stability of the bifurcating periodic
orbits: the bifurcating periodic orbits is stable (unstable) if
$K_2<0\,(>0)$.

\section{A numerical example}
Consider an example in the form of system (1), with
$a=1,\,b=1.1,\,c=1,\,\tau=1, \mbox{ and } f(.)=\mbox{tanh}(.)$.
The theoretical analysis in Section 2 leads to
\begin{equation*}
\omega_0=0.9017, \hspace{1cm} h_0=1.1496
\end{equation*}
It then follows from the results in Section 3 that
\begin{equation*}
K_1=0.2627, \hspace{1cm} K_2=-0.3408
\end{equation*}
These calculations prove that the equilibrium (0, 0) is stable
when $h\, < \, h_0$, as shown by Fig. 1, where $h=1.1$. When $h$
passes through the critical value $h_0=1.1496$, the equilibrium
losses its stability and a Hopf bifurcation occurs, i.e., a family
of periodic solutions bifurcate from the equilibrium. Each
individual periodic orbit is stable for $K_2<0$. Since $K_1K_2<0$,
the bifurcating periodic solutions exist at least for values of
$h$ slightly larger than the critical value. Choosing $h=1.4$, as
predicted by the theory, Fig. 2 shows that there is an orbitally
stable limit cycle.
\begin{figure}[htb]
\centering
\includegraphics[width=15cm]{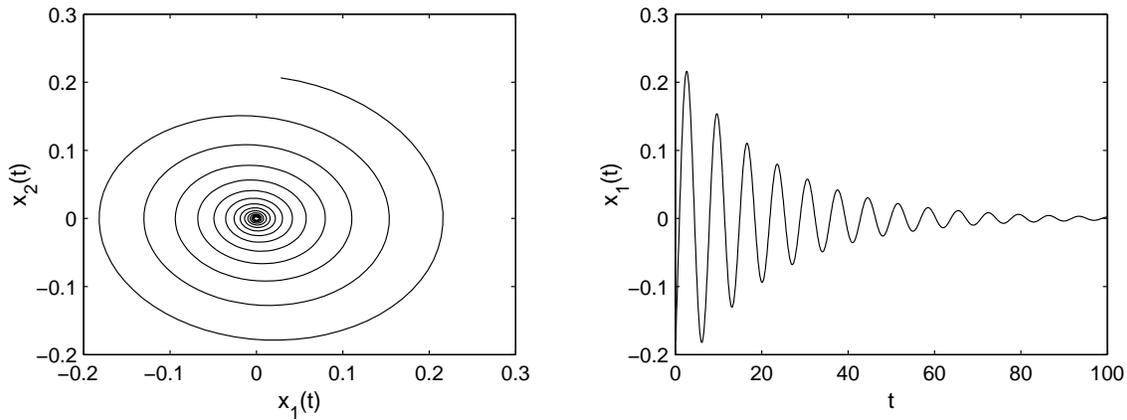}
\caption{Phase plot and waveform plot for system (1) with $h=1.1$}
\end{figure}
\begin{figure}[htb]
\centering
\includegraphics[width=15cm]{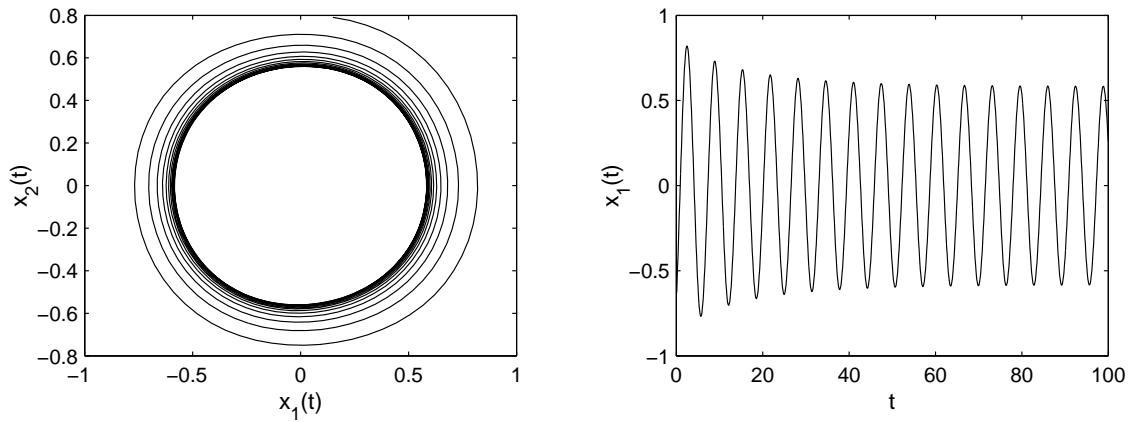}
\caption{Phase plot and waveform plot for system (1) with $h=1.4$}
\end{figure}

\section{Chaotic behavior}
In this section, we study the dynamical behavior of system (1)
with the activation function $f(x)=xe^{-x^2/2}$ and
$a=0.8,b=1,c=5,\tau=5$, and let $h$ be a variable parameter. It is
noted that several other parameters have also been examined and
shown to exhibit similar dynamical phenomena. Due to limitation of
space, those results are not presented here.

When $h<0.65$, the system is stable. When increasing $h$ to
$h=0.7$, the system produces a periodic orbit. When $h=0.7$, the
phase plot and the waveform plot of $x_1(t)$ is shown in Fig. 3.
When the value of $h$ passes 0.9, the system becomes chaotic. In
Fig. 4, we show the phase plot when $h=1.0$, and in Fig. 5 we show
the waveform of $x_1(t)$ and the power spectrum plots when
$h=1.0$. When $h=5.0$, the phase plot and the waveform of
$x_1(t)$, and the power spectrum plots, are shown in Fig. 6 and
Fig. 7, respectively. From these figures, we can see that the
system is also chaotic, but it is different from the case of
$h=1.0$.
\begin{figure}[htb]
\centering
\includegraphics[width=15cm]{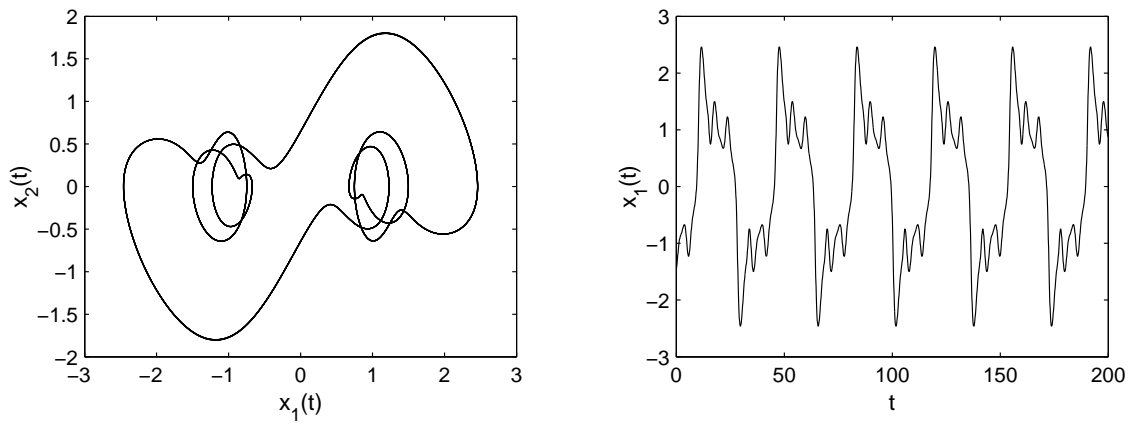}
\caption{Phase plot and waveform plot when $h=0.7$}
\end{figure}
\begin{figure}[htb]
\centering
\includegraphics[width=10cm]{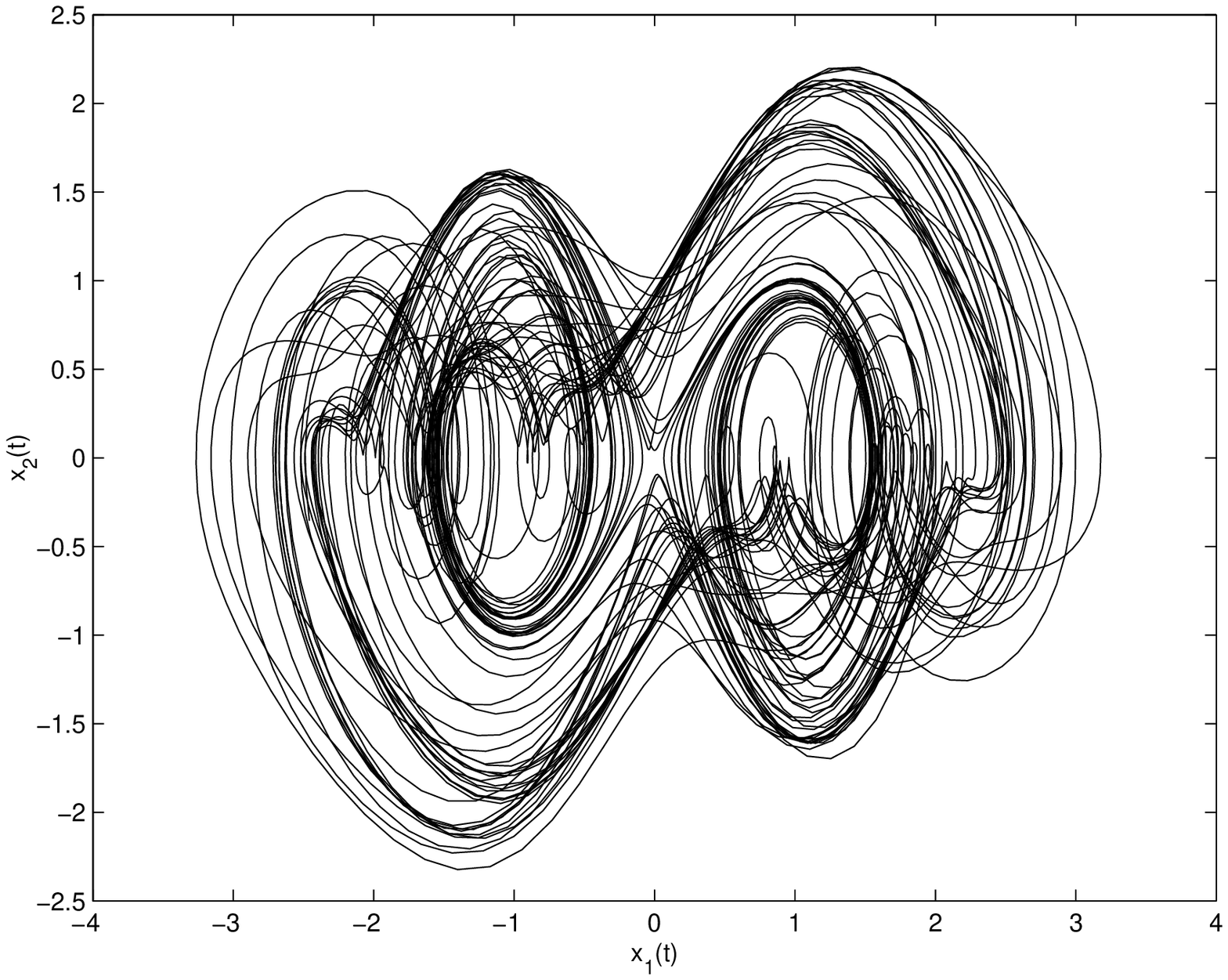}
\caption{Phase plot when $h=1.0$}
\end{figure}
\begin{figure}[htb]
\centering
\includegraphics[width=15cm]{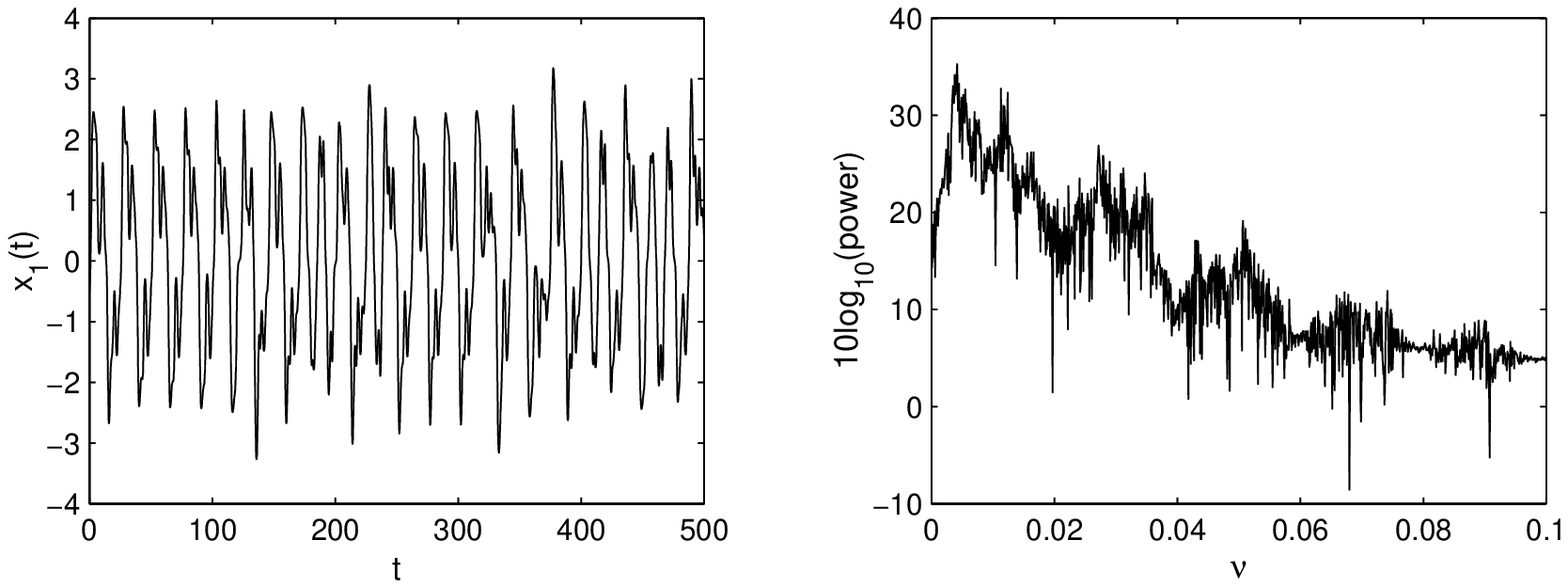}
\caption{Waveform and power spectrum plots when $h=1.0$}
\end{figure}
\begin{figure}[htb]
\centering
\includegraphics[width=10cm]{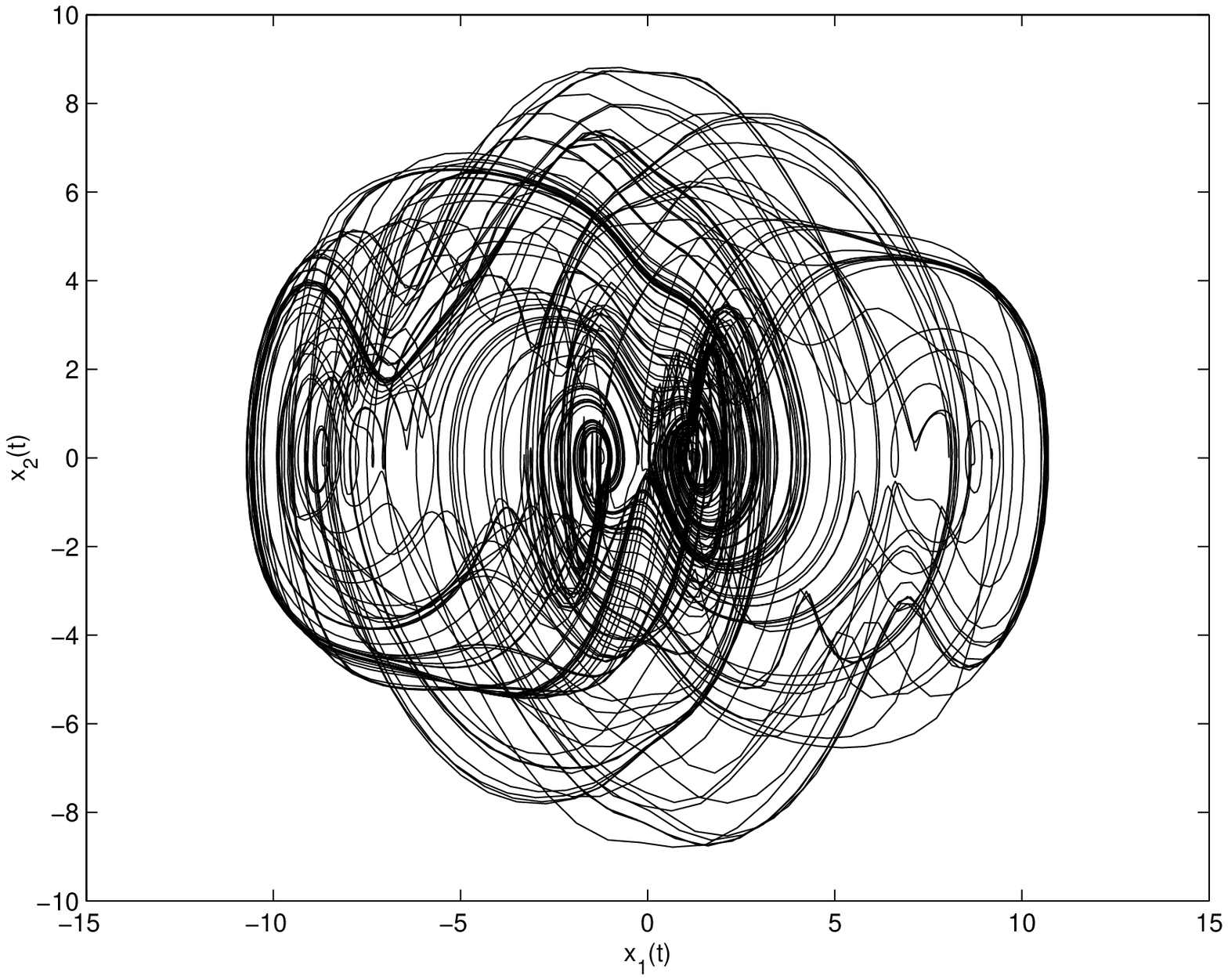}
\caption{Phase plot when $h=5.0$}
\end{figure}
\begin{figure}[htb]
\centering
\includegraphics[width=15cm]{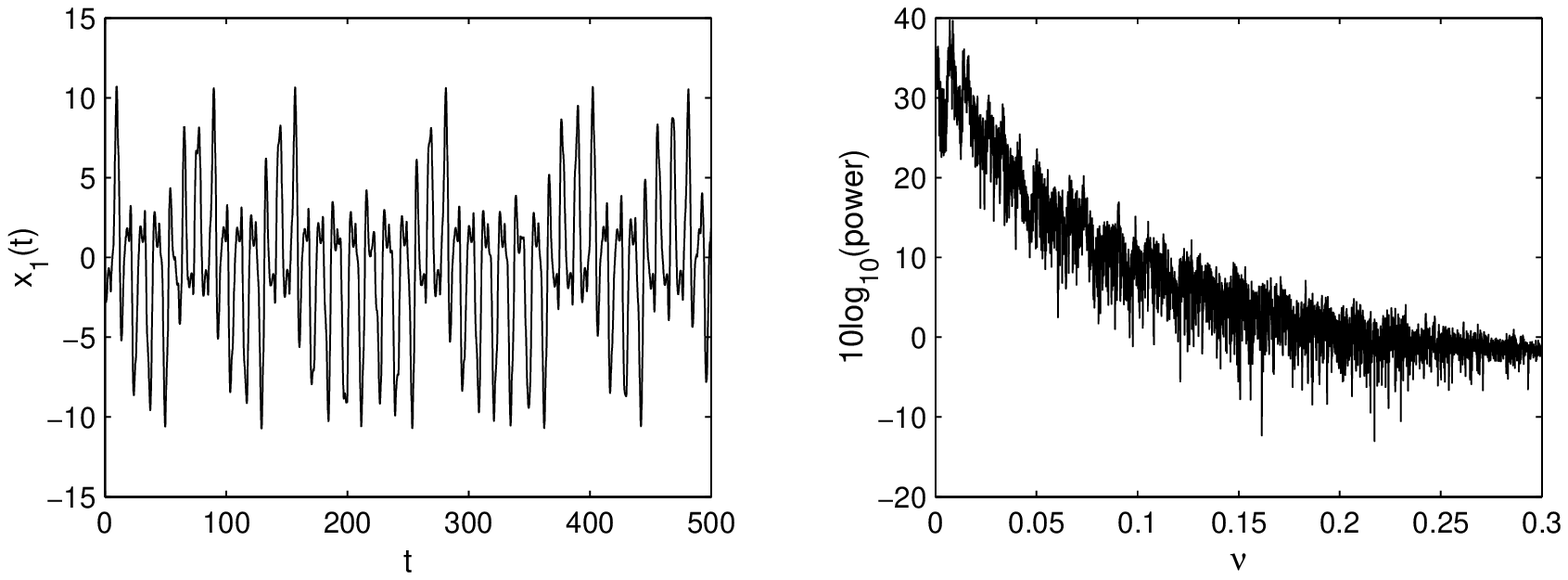}
\caption{Waveform and power spectrum plots when $h=5.0$}
\end{figure}

\section{Conclusions}
A single delayed neuron model with inertial term has been
investigated in this paper. Hopf bifurcation is studied by using
the normal form theory of retarded FDEs, in which the coefficients
of the normal form are obtained in terms of the original delayed
equation directly, without the need to compute the center manifold
beforehand, which simplifies the computation procedure. With a
nonmonotonic activation function, chaotic behavior has also been
observed in this system.

\section*{Acknowledgments}
This work was supported by the National Natural Science Foundation
of China under Grant 60271019 and the Hong Kong Research Grants
Council under the CERG grant CityU 1004/02E.

\end{document}